\documentclass[reprint,amsmath,amssymb,aps,prb,twocolumn,superscriptaddress]{revtex4-1}
\usepackage{lipsum}
\usepackage{graphicx}
\usepackage{dcolumn}
\usepackage{bm}
\usepackage{xcolor}

\begin{document}
\preprint{APS/123-QED}

\title{Magnetization switching in the inertial regime}

\author{Kumar Neeraj}
\affiliation{Department of Physics, Stockholm University, 106 91 Stockholm, Sweden}

\author{Matteo Pancaldi}
\altaffiliation[Current address: ]{Elettra-Sincrotrone Trieste S.C.p.A., 34149 Basovizza, Trieste, Italy}
\affiliation{Department of Physics, Stockholm University, 106 91 Stockholm, Sweden}

\author{Valentino Scalera}
\affiliation{Faculty of Computer Science, Free University of Bozen-Bolzano, Bolzano, Italy}

\author{Salvatore Perna}
\affiliation{DIETI, University of Naples Federico II, Naples, Italy}

\author{Massimiliano d’Aquino}
\affiliation{DIETI, University of Naples Federico II, Naples, Italy}

\author{Claudio Serpico}
\affiliation{DIETI, University of Naples Federico II, Naples, Italy}

\author{Stefano Bonetti}
\email[Corresponding author: ]{stefano.bonetti@fysik.su.se}
\affiliation{Department of Physics, Stockholm University, 106 91 Stockholm, Sweden}
\affiliation{Department of Molecular Sciences and Nanosystems, Ca’ Foscari University of Venice, 30172 Venezia-Mestre, Italy}

\begin{abstract}
We have numerically solved the Landau-Lifshitz-Gilbert (LLG) equation in its standard and inertial forms to study the magnetization switching dynamics in a $3d$ thin film ferromagnet. The dynamics is triggered by ultrashort magnetic field pulses of varying width and amplitude in the picosecond and Tesla range. We have compared the solutions of the two equations in terms of switching characteristic, speed and energy analysis. Both equations return qualitatively similar switching dynamics, characterized by regions of slower \emph{precessional} behavior and faster \emph{ballistic} motion. In case of inertial dynamics, ballistic switching is found in a 25\% wider region in the parameter space given by the magnetic field amplitude and width. The energy analysis of the dynamics is qualitatively different for the standard and inertial LLG equations. In the latter case, an extra energy channel, interpreted as the kinetic energy of the system, is available. Such extra channel is responsible for a resonant energy absorption at THz frequencies, consistent with the occurence of spin nutation.
\end{abstract}

\maketitle

\section{\label{sec:level1}Introduction}
The traditional method of writing information in magnetic hard disk drives consists in reversing the magnetization direction via the application of magnetic fields produced by external currents and localized via a so-called "write-head". In order to achieve efficient switching, the magnetic field is applied nearly anti-parallel to the direction of the initial magnetization state, and the switching process thus obtain is often referred to as ``damped'' switching \cite{mallinson_damped_2000,bertotti2003comparison, stohr2006magnetism}. The switching time in this process is limited by the macroscopic relaxation time of the magnetization of the order of 100 ps, and correctly described by the standard Landau Lifhitz Gilbert (LLG) equation \cite{gilbert2004phenomenological}. Geritts \textit{et. al.} \cite{gerrits2002ultrafast} demonstrated instead a technique by which ultrafast magnetization reversal can be achieved using picosecond-long magnetic field pulses \emph{transverse} to the magnetization direction. Such a switching technique was also reported in other studies \cite{kaka2002precessional,hiebert2002comparison}. The switching times reported in these works depend on the amplitude of the magnetic field and on the duration of the pulse, with a general trend that a pulse of larger amplitude will reduce the switching time. Coincidentally, at the same time, a series of experiments performed at the Stanford linear accelerator demonstrated the ultrafast switching of magnetization by intense magnetic fields created by relativistic electron bunches. In the paper published by Tudosa \textit{et al.} \cite{tudosa2004ultimate}, it was argued that deterministic magnetization switching cannot occur faster than 2 ps, setting this as the ultimate speed for magnetic reversal. However, due to the complexity of such an accelerator based experiment, direct observation of such an ultrafast switching in the time domain was not possible.

A few years later, it was shown that a novel idea, i.e. using magnetic inertia, can greatly enhance the switching speed in antiferromagnets \cite{kimel2009inertia}, up to 10 times faster than that reported in the above-mentioned studies. So far, nutation-type magnetization motions in ferromagnetic systems have been studied mostly theoretically, in the framework of classical LLG dynamics at GHz frequencies \cite{serpico_quasiperiodic_2004}, as well as in other numerical studies considering the inertial version of the LLG, often refered to as iLLG equation \cite{bottcher2011atomistic,bhattacharjee2012atomistic,bastardis2018magnetization, makhfudz2020nutation}. These works predicted the appearance of a spin nutation with an intrinsic resonance in the $10^{11}-10^{15}$ Hz range. The direct detection of spin nutation in ferromagnets recently achieved experimentally \cite{neeraj2021inertial} has allowed to narrow down the rather broad frequency range to the $10^{12}$ Hz one, i.e. in the THz region, for typical $3d$ ferromagnetic alloys such as NiFe and CoFeB.

In this work, we explore the role of inertia in the magnetization switching of a thin film ferromagnet triggered by magnetic field pulses in the picosecond range. Using macrospin simulations, we create a map of the magnetization dynamics as a function of the pulse duration and amplitude. The dynamics is obtained and analyzed solving both the standard LLG equation and its inertial form, the so-called iLLG equation. We used realistic material parameters for an archetypal ferromagnet, namely Ni$_{81}$Fe$_{19}$ (permalloy), but they are valid for all thin film ferromagnets with similar characteristics. We show that the iLLG simulations predict a larger stability region than the LLG one for one the two characteristic switching processes, and that the effect of the nutation resonance can be detected in switching experiments. We discuss our results analyzing the dynamics of the classical energy terms.

\section{\label{sec:level2}Methods}
\begin{figure*}[t]
\centering
\includegraphics[width=\textwidth]{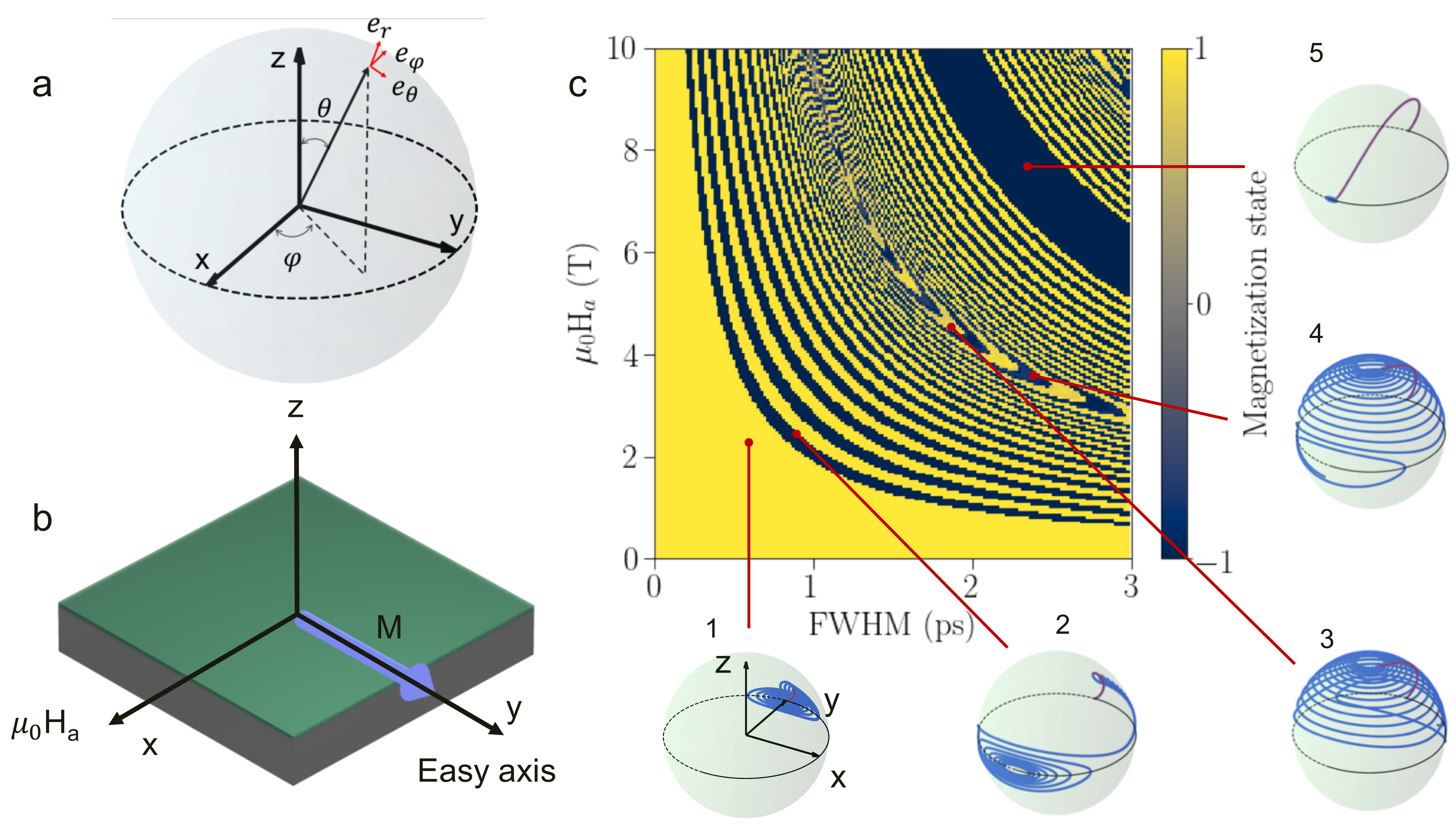}
\caption{(a) Spherical coordinates used for the numerical simulations described in the main text. (b) Geometry of the thin film system considered. The easy magnetization axis lies along the \textit{y}-direction, and the in-plane hard magnetization axis is along the \textit{x}-direction. (c) Main plot: magnetization state diagram for different magnetic pulse amplitude and FWHM. Side plots: magnetization precession trajectories in selected points of the diagram. The color bar shows the \textit{y} component of the magnetization vector at the end of the simulation. The magnetization starts always from the positive \textit{y} direction, i.e. aligned parallel to the easy magnetization axis.}
\label{fig_1}
\end{figure*}
Until recently, the dynamics of magnetization switching in ferromagnetic systems has been mostly described by the conventional LLG equation \cite{gilbert2004phenomenological}, which can be written as 
\begin{equation} \label{eq1}
   \frac{d{\textbf{M}}}{dt} = -|\gamma| {\textbf{M}} \times \left({\textbf{H}_\textrm{eff}} - \frac{\alpha}{|\gamma| M_{s}}\frac{d\textbf{M}}{dt}\right),
\end{equation}
where $|\gamma|/(\mu_0 2\pi) \approx 28 $ GHz/T is the gyromagnetic ratio, $\textbf{H}_\textrm{eff}$ is the effective magnetic field, calculated as the variational derivative of energy with respect to the magnetization, $M_{s}$ is the saturation magnetization, and $\alpha$ is the Gilbert damping. The first term on the right-hand side describes the precession motion while the second describes the damping of precession \cite{gilbert2004phenomenological}. However, according to recent theoretical studies the inclusion of an extra inertial term has been suggested \cite{ciornei2011magnetization,olive2012beyond, fahnle2011generalized, mondal2017relativistic} as
\begin{equation} \label{eq2}
\frac{d{\textbf{M}}}{dt} = -|\gamma| {\textbf{M}} \times \left[{\textbf{H}_\textrm{eff}} - \frac{\alpha}{|\gamma| M_{s}}\left(\frac{d{\textbf{M}}}{dt} + \tau \frac{d^2 {\textbf{M}}}{dt^2}\right)\right],
\end{equation}
where $\tau$ is the angular momentum relaxation time. The extra term is the second derivative of the magnetization vector which leads to spin nutation, in addition to precession and damping. The microscopic origin of inertia is still not clearly understood and several theoretical studies have been presented  \cite{mondal2017relativistic,makhfudz2020nutation}. Mondal \textit{et al} \cite{mondal2017relativistic} have suggested that the inclusion of higher order spin-orbit coupling terms in the Hamiltonian of the system will produce an equation of this kind.

In order to understand the dynamical behavior of magnetization shown by Eqs. (\ref{eq1}) and (\ref{eq2}) we perform simulations on an infinite thin magnetic film, assuming that the magnetization is spatially homogeneous (macrospin approximation). The effective magnetic field $\textbf{H}_\textrm{eff}$ appearing in Eq. \eqref{eq2} takes into account the different interactions occurring among elementary magnetic moments, namely
\begin{equation} \label{heff}
\textbf{H}_\textrm{eff} = \textbf{H}_\textrm{a}+\textbf{H}_\textrm{ani}+\textbf{H}_\textrm{m},
\end{equation}
where $\textbf{H}_\textrm{a}$ is the applied magnetic field, $\textbf{H}_\textrm{ani} = 2K_1 / (\mu_0 M_s)\bm e_y$ is the uniaxial magneto-crystalline anisotropy field with $K_1$ being the anisotropy constant and $\textbf{H}_\textrm{m} = - {N}\cdot\bm M$ the magnetostatic (demagnetizing) field. $N$ is the demagnetization tensor, with ${N = \text{diag}(N_x,N_y,N_z)}$ when referred to the principal axes of the system. For the case of a thin film infinite in the $x,y$ directions, the demagnetizing field takes the form $\textbf{H}_\textrm{m} = - M_z\bm e_z$ \cite{devolder2006precessional,d2004nonlinear,bauer2000switching}. We consider a thin film with an in-plane easy and hard axis, with an anisotropy field of $\mu_0\textbf{H}_\textrm{ani} = 0.1$ T, as shown in Fig. \ref{fig_1}(b). This anisotropy field defines two equilibrium states of magnetization in the system along the \textit{y}-axis.

We solve the LLG and iLLG differential equations using two independent numerical codes. The first one is based on a classical Runge-Kutta fourth-order method in spherical coordinates defined by a set of three unit vectors \{$\textbf{e}_\textrm{x}, \textbf{e}_\textrm{y}, \textbf{e}_\textrm{y}$\} as shown in Fig. \ref{fig_1}(a). Details are given in Appendix A. The second code relies on the appropriate extension to the iLLG dynamics of the implicit midpoint rule time-stepping \cite{daquino_numerical_2005} in Cartesian coordinates. The considered magnetic thin film has material parameters similar to those of polycrystalline permalloy, i.e. $\alpha = 0.023$, $\mu_0M_s = 0.92$ T, $\tau = 11.3$ ps \cite{neeraj2021inertial}. The applied magnetic field has a Gaussian-like shape with a Full-Width Half-Maximum (FWHM) varying from 0 to 3 ps in steps of 10 fs, while the amplitude ($\mu_0\textbf{H}_\textrm{a}$) is varied in steps of 100 mT from 0 to 10 T. The field is always applied along the $x$-axis, i.e. perpendicular to the easy magnetization axis and equilibrium states.

\section{\label{sec:level3}Results and discussion}
For a thin film system, magnetization switching can be understood as a three step process \cite{stohr2006magnetism}. (\emph{i}) An applied pulse \textbf{H}$_\textrm{a}$ lying in the film plane and perpendicular to \textbf{M} will exert a Zeeman torque, and will lead \textbf{M} to precess out of the in-plane easy axis. (\emph{ii}) This torque will simultaneously create a demagnetizing field perpendicular to the film plane. The demagnetizing field will further lead to precession of \textbf{M} around the axis perpendicular to the film plane. (\emph{iii}) Eventually, once the applied magnetic field is turned off, the magnetization relaxes along the direction of the effective magnetic field defined in Eq.~\eqref{heff}.

Figure \ref{fig_1}(c) shows the diagram of the magnetization switching obtained by numerically solving the iLLG equation. Similar diagrams have been extensively used for studying magnetization switching in different case studies \cite{bertotti_geometrical_2003,bhattacharjee2012theoretical,wienholdt2012thz,shutyi2020multistability}, but not yet for the case of inertial dynamics in ferromagnets. The diagram shows the state of magnetization with respect to the applied Gaussian pulse width and amplitude, with the yellow and blue regions representing the non-switched and, respectively, the switched state of magnetization. It should be noted that the presence
of an in-plane hard axis creates an energy landscape characterized by two stable states along the $y$-axis, two saddle-type equilibria along the $x$-axis, and two unstable equilibria along the $z$-axis.

In all the regions of the diagram, we can extract the full magnetization dynamics, as shown in the trajectories on the unitary sphere in Fig.~\ref{fig_1}(c) for a few characteristics cases. Case 1 is representative of the region where the magnetization state after excitation relaxes back to its initial equilibrium state. In case 2, the applied pulse deposits enough energy into the system for the magnetization to cross the barrier created by the presence of the in-plane hard axis, and to switch to the opposite magnetization state. We denote this type of switching as the \textit{precessional} switching. For cases 3 and 4, the magnetization reaches just close to the energy barrier (i.e., the top of the sphere) of the out-of-plane hard axis where it precesses several times before relaxing to the initial and, respectively opposite state. It is worthwhile to point out that this type of switching exhibits quasi-random relaxation behavior as a result of the multi-stability and the low dissipation of magnetization dynamics\cite{serpico_analytical_2009}. The most peculiar region is the one represented by case 5, where the magnetization is able to cross the energy barrier of the out-of-plane hard axis and switches to the opposite magnetization state with negligible precession. We denote this type of switching as \textit{ballistic} switching. Note that in literature the terms \emph{precessional} and \emph{ballistic} are sometimes used interchangeably \cite{daquino_numerical_2004,nozaki2006numerical,bazaliy2011analytic}. A reason for this situation is the fact that previous studies considered only slower time scales, where switching such as the one in case 5 was never observed. Here, considering much faster dynamics, we are able to identify the qualitative difference between the two switching processes.

\begin{figure}[t]
\includegraphics[width=\columnwidth]{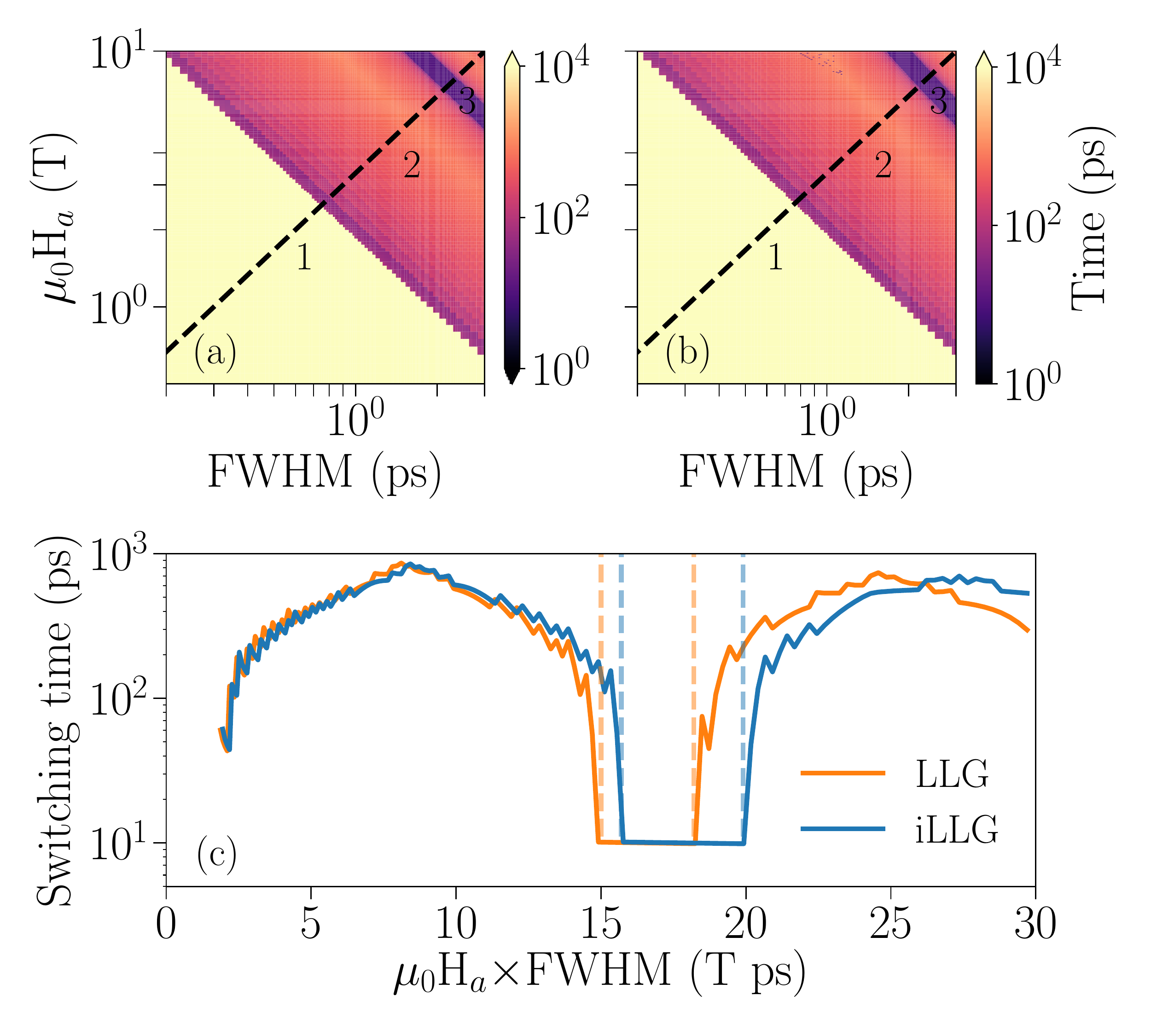}
\caption{\label{fig_2} Magnetization switching times calculated for different pulse amplitude and FWHM width using (a) the LLG and (b) the iLLG equations. All axes and amplitudes are in logarithmic scale. (c) Switching time along the diagonal line cuts (i.e. perpendicular to the lines of constant pulse energy) shown by the white solid lines in (a) and (b). The dashed vertical lines indicate the boundaries of the ballistic switching region for the two simulations.}
\end{figure}

From the dynamical response of the magnetization to different Gaussian pulses, the magnetization switching time can be obtained. The exact definition of switching time reported in literature is ambiguous and no convention is reported \cite{stohr2006magnetism}. In some studies the switching time is defined to be the time taken by the magnetization to completely reach equilibrium along the opposite easy axis direction. In this study, we define the magnetization switching time as the time taken for magnetization to cross the  energy barrier for the last time before relaxing on either side of the easy axis (+\textit{y} or -\textit{y} axis in our geometry). Based on this definition (see Appendix B for the details) we calculate a diagram of the magnetization switching time for different FWHM and amplitude of the externally applied magnetic field pulse as shown in Fig. \ref{fig_2}. Fig. \ref{fig_2}(a) and (b) were obtained by solving the LLG and, respectively, the iLLG equation.

In both plots, the initial black regions (marked as 1) correspond to the regions where the magnetization never crosses the energy barrier, and therefore there is no switching. On the other hand, region 2 indicates the area of relatively longer switching times, where the magnetization precesses several times around the unitary magnetization sphere before relaxing to either one of the two equilibrium states. Once again, as in Fig. \ref{fig_1}, the most peculiar observation is found in region 3, which corresponds to the case of ballistic switching. Here, we notice that the switching time is the fastest and is largely independent of the magnetic pulse width and amplitude. We highlight the choice of a logarithmic scale to properly highlight the different orders of magnitude between the switching times in regions 2 and 3. A closer look at the difference in the two simulation results shows that the width of region 3 for the case of the iLLG dynamics is larger than that of the LLG one, as highlighted by the dashed vertical lines in Fig.~\ref{fig_2}(c). The figure is obtained taking a diagonal cut of Fig.~\ref{fig_2}(a),(b). The width of the ballistic switching region is approximately 3.34 T ps for the LLG dynamics, and 4.16 T ps for the iLLG one, i.e. approximately 25\% larger in the latter case. We also notice that, for the LLG dynamics, the ballistic region starts at lower $\mu_0$H$_\textrm{a}\times$FWHM values than for the iLLG case.

We can interpret these observations with analogies taken from classical mechanics. For a given driving force, a system at rest with larger inertia will react less to that force, but then inertia helps in preserving the motion of the system once the dynamics has started. 

\begin{figure}[t]
\centering
\includegraphics[width=\columnwidth]{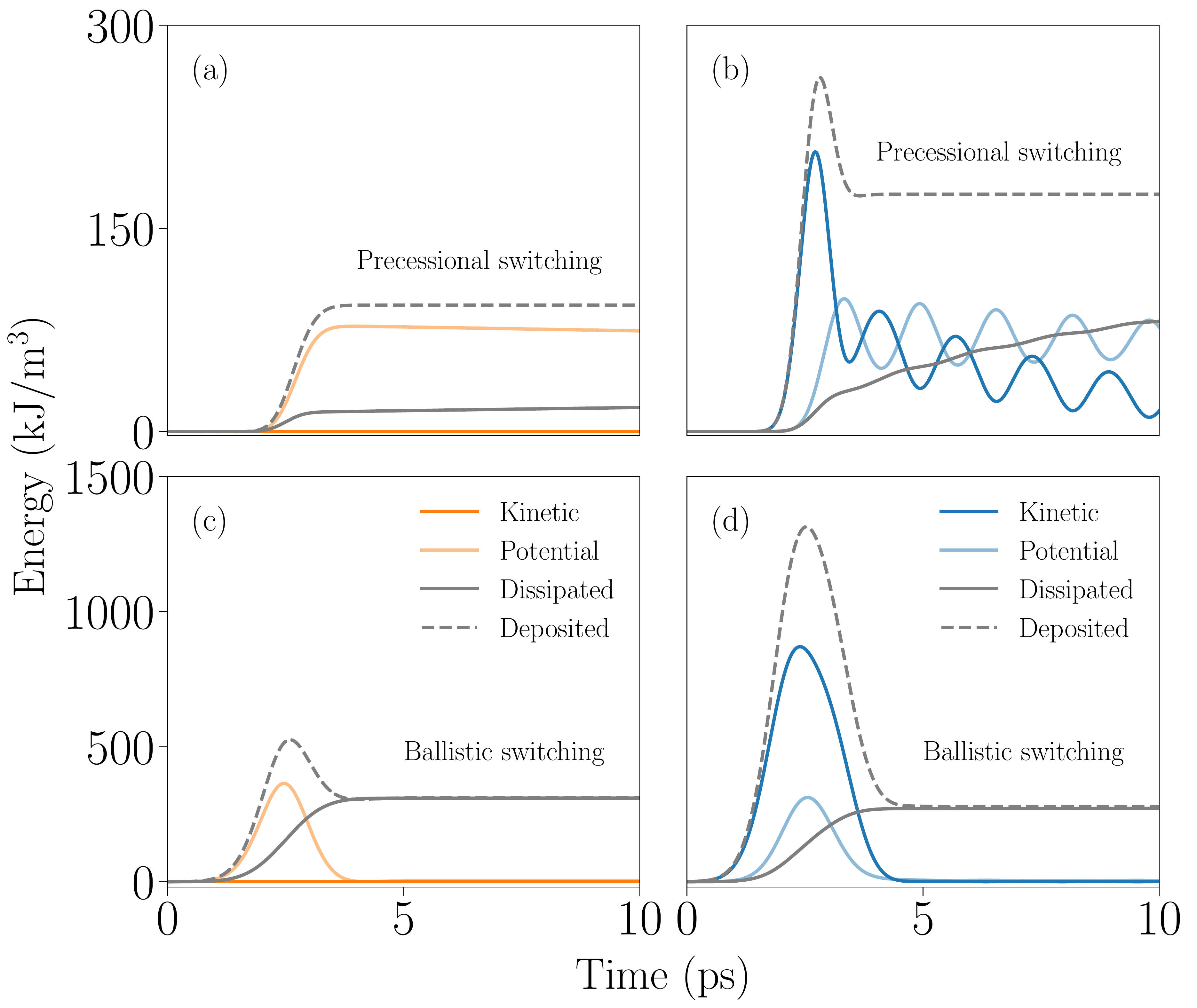}
\caption{Temporal evolution of the kinetic, potential, dissipated and deposited energy terms for a few selected simulation parameters. (a) LLG and (b) iLLG dynamics of the different energy terms for an applied magnetic field of 2 T amplitude and 1 ps FWHM, i.e. in the \emph{precessional} switching region. (c) LLG and (d) iLLG energy dynamics for 8 T, 2.1 ps magnetic field pulse, i.e. in the \emph{ballistic} switching region.}
\label{fig_3}
\end{figure}

In addition to the speed of the switching process, it is equally important to understand the energy associated with it. In order to do this, we calculated the work per unit volume $\Delta W$ performed by the field pulse on the magnetic system as
\begin{equation} \label{en}
\Delta W = \int_0^{t_p} \mu_0\textbf{H}_\textrm{a} \cdot \frac{d\textbf{M}}{dt}dt =  \frac{\mu_0}{\gamma  M_s} \int_0^\infty \alpha \bigg|\frac{d\bm M}{dt}\bigg|^2 \,dt \,,
\end{equation}
where $t_p$ is the time instant at which the external pulse amplitude goes back to zero after reaching its maximum. The latter formula comes from the energy balance equation (see the derivation in Appendix C), i.e.
\begin{equation} \label{enbal}
\int_0^{t} \mu_0 \textbf{H}_\textrm{a} \cdot \frac{d\textbf{M}}{dt}dt =   \int_0^t \bigg( \frac{dA}{dt} + \frac{dK}{dt}+\frac{\alpha\, \mu_0}{\gamma M_s} \bigg|\frac{d \bm M}{dt}\bigg|^2\bigg)dt \,.
\end{equation}

The left-hand side of Eq. (\ref{enbal}) is the energy deposited by the pulse as function of time $t$ after turning the field on. This is  equal to the energy density absorbed by the system, which in turn is the sum of the potential energy $A = K_1 (1-(M_y/M_s)^2)+ \dfrac{1}{2}\mu_0 M_z^2$, kinetic energy $K = \dfrac{\mu_0}{2\gamma M_s} \alpha\tau\displaystyle \left|\frac{d\textbf{M}}{dt}\right|^2$, and dissipated energy $\displaystyle\int_0^t \dfrac{\mu_0}{\gamma M_s}\alpha \left|\dfrac{d\textbf{M}}{dt}\right|^2 \,dt$. Eq.~(\ref{enbal}) is derived for the iLLG case but it is directly applicable to LLG equation without the kinetic term, since $\tau=0$ in this case.

Using Eq.~\eqref{enbal}, we plot in Fig. \ref{fig_3} the temporal evolution of the energy transfer from the magnetic field pulse into the system without (LLG) and with inertia (iLLG), and for the case of precessional (magnetic field with amplitude 2 T, and 1 ps FWHM duration) and ballistic switching (8 T, 2.1 ps FWHM). In all cases, at long enough times, the dissipated energy converges to the value of the deposited energy, indicated with solid and, respectively, dashed black lines. However, the way such values is reached is remarkably different for the four cases.

In the case of precessional switching in the standard LLG dynamics, Fig. \ref{fig_3}(a), the energy is dissipated rather slowly, consistent with the comparatively long switching times shown in Fig. \ref{fig_2}. In the energy picture, the potential energy of the precession is dissipated away by the Gilbert damping in the hundreds of ps range. For the precessional switching in the case of iLLG dynamics, Fig. \ref{fig_3}(b,) the dissipation of the deposited energy is again a slow process, but there are two key differences with the LLG case. First, the total deposited energy is approximately twice as large, owing to the fact that an additional energy channel, i.e. the kinetic one, is now available, which also creates a fast overshoot at the beginning of the dynamics. Second, in addition to the slow dissipation of energy, there is now a relatively fast periodic exchange between potential and kinetic energies. This takes place at THz rates and it is the signature of spin nutation in the energy dynamics. Interestingly, the kinetic energy relaxes faster than the potential energy.

\begin{figure}[t]
\includegraphics[width=\columnwidth]{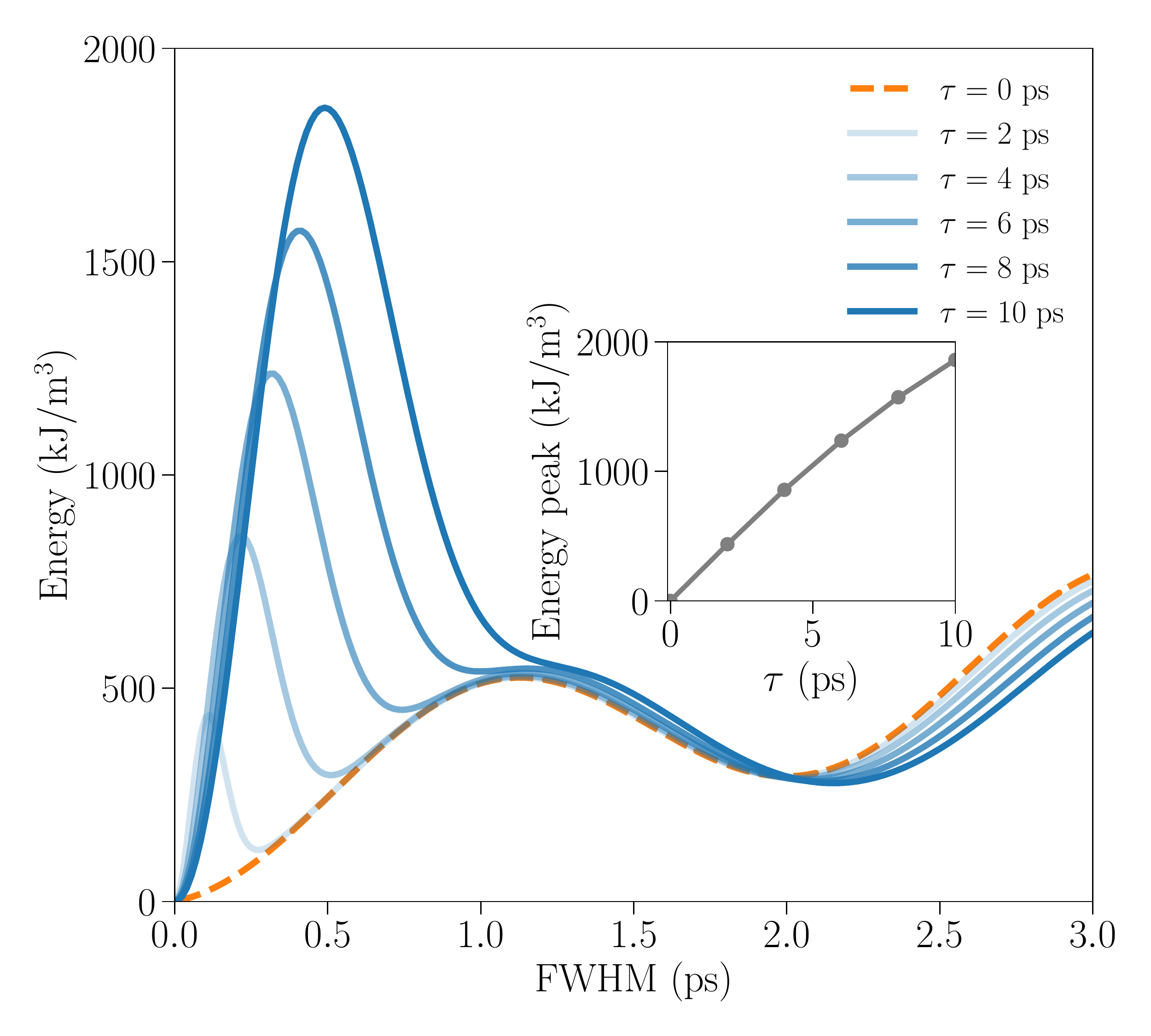}
\caption{Total energy deposited by a 8 T magnetic field pulse as a function of pulse FWHM. The plot is computed for different values of the angular momentum relaxation time constant $\tau$ used in the iLLG dynamics. Inset: amplitude of peak energy absorbed as a function of $\tau$.}
\label{fig_4}
\end{figure}

We now turn to the case of the ballistic switching, i.e. Figs. \ref{fig_3}(c),(d). Here, the dissipated energy for the LLG and iLLG dynamics is almost identical, but in the case of the iLLG dynamics in Fig. \ref{fig_3}(d) there is, similarly to Fig. \ref{fig_3}(b), a larger overshoot than for the LLG case. This is again due to the additional contribution of the kinetic energy term. This overshoot is intriguing and it is a rather general feature of the iLLG dynamics. We demonstrate this by calculating the LLG and iLLG diagrams for all magnetic field amplitudes and widths considered in Figs. \ref{fig_1} and \ref{fig_2}, and we plot this in Fig. \ref{fig_6} in Appendix C. In order to investigate this observation further, we take a horizontal line cut in Fig. \ref{fig_6}(b) in correspondence of the 8 T value, and we plot it in Fig. \ref{fig_4}. A peak is present in the deposited energy value for a pulse FWHM of around 0.5 ps and $\tau=10$ ps. By varying the value of $\tau$, we observe that the maximum work is performed by the applied field when the pulse duration 6$\sigma$, with $\sigma\approx\text{FWHM}/2.4$, approximately matches  the period of the nutation resonance $2\pi \alpha \tau$. Hence, we can attribute the occurrence of the peak to the resonant absorption of the nutation resonance. In the inset of Fig. \ref{fig_4} we also plot the peak energy value for the different $\tau$, observing a monotonic increase. This is generally consistent with a torque-driven dynamics with a constant gyromagnetic ratio $\gamma$ where, for a given magnetic field amplitude, the larger torque on the magnetization, hence its maximum displacement, is achieved when the magnetic field varies more slowly.

\section{\label{sec:level4}Summary and conclusion}
From our LLG and iLLG simulations we explored the different switching dynamics triggered by ultrafast magnetic field pulses of different FWHM and amplitude in the ps and Tesla ranges. Depending on those parameters, the magnetization exhibits \textit{precessional} or \textit{ballistic} switching.The ballistic switching is always much faster than the precessional switching, for both the LLG and iLLG equations. We also observed that the width of the ballistic switching region for the iLLG case is 25\% larger than the LLG one. From an application perspective, this difference in width of the ballistic region may turn out to be useful for the reliability of ultrafast switching, and guide the choice of materials for magnetic storage devices with larger inertia.

We further showed how the energy is deposited into the system by the external magnetic field pulse. For the iLLG dynamics, the external pulse drives the spin nutation with a characteristic resonant feature, and the deposited energy has an additional kinetic energy channel not available in the standard LLG dynamics, where instead only the potential energy channel exists.

Our results will be useful for the design of devices relying on ultrafast magnetization switching using picosecond magnetic field pulses. The realization of such short and intense magnetic fields have been suggested recently designing magnetic-field enhancing metamaterials \cite{polley2018thz}, by the use of vector laser beams \cite{sederberg2020tesla} or thanks to ultrafast electronic switches \cite{jhuria2020spin}. Finally, we anticipate that our findings will be relevant to much of the recent works in the magnetism community aimed at the understanding of inertial spin dynamics \cite{thonig2017magnetic,giordano2020derivation,mondal2021nutation,mondal2021theroy,titov2021inertial,thibaudeau2021emerging,lomonosov2021anatomy,ruggeri2021numerical,cherkasskii2021dispersion,rahman2021observable,titov2021deterministic,anders2020versatile,lou2021large,gupta2021co2feal}.

\begin{acknowledgments}
K.N., M.P. and S.B. acknowledge support from the European Research Council, Starting Grant 715452 MAGNETIC-SPEED-LIMIT.
\end{acknowledgments}

\appendix
\section{\label{sec:level5}Magnetization dynamics in spherical coordinates}
The dynamical Eqs. \eqref{eq1} and \eqref{eq2} can be decomposed in the reference frame of a spherical coordinate system defined by a set of three unit vectors \{$\textbf{e}_\textrm{r},\textbf{e}_\mathrm{\theta},\textbf{e}_\mathrm{\phi}$\}. The magnetization vector in this reference frame is defined by the set of coordinates $M = (M_{s}, \theta,\phi )$, where the unit vector  $\textbf{e}_\textrm{r}$ is aligned with $\textbf{M}$ while \textbf{e}$_\mathrm{\theta}$ and $\textbf{e}_\mathrm{\phi}$ point in the direction of increasing $\theta$ and $\phi$ respectively. They are both tangent to the unit sphere, as shown in Fig \ref{fig_1}(a). The expression for the net magnetic field in angular components can be written in terms of the Cartesian coordinates  as 
\noindent

\begin{align} \label{eq3}
\textbf{H}_\textrm{r} &= \textbf{H}_\textrm{x}\cos \phi \sin \theta+\textbf{H}_\textrm{y}+\sin \phi \cos \theta + \textbf{H}_\textrm{z} \cos \theta\\
\textbf{H}_\mathrm{\theta} &= \textbf{H}_\textrm{x}\cos \phi \sin \theta+\textbf{H}_\textrm{y}+\sin \phi \cos \theta - \textbf{H}_\textrm{z} \sin \theta\\
\textbf{H}_\mathrm{\phi} &= -\textbf{H}_\textrm{x} \sin \phi+\textbf{H}_\textrm{y}\cos \phi \,.
\end{align}

\subsection{LLG equation in spherical coordinates}
Since $\displaystyle{\frac{d\mathbf{M}}{dt}}$ is perpendicular to \textbf{M}, a vector multiplication of Eq. \eqref{eq1} by \textbf{M} will give
\begin{equation} \label{eq4}
    (-\gamma \eta M_\textrm{s}^2 + \textbf{M} \times)  \frac{d\mathbf{M}}{dt}= -M_{s}^{2}\textbf{H}_\mathrm{eff \perp},
\end{equation}
\noindent 
where $\textbf{H}_\mathrm{eff \perp}$ is the component of effect field perpendicular to \textbf{M}. The time derivative of the magnetization in spherical coordinates is written as, $\displaystyle\frac{d\mathbf{M}}{dt} = M_s \frac{d\theta}{dt}  \textbf{e} _\mathrm{\theta} + M_s \frac{d\phi}{dt} \sin \theta \textbf{e}_\mathrm{\phi}$. Plugging this into Eq. \eqref{eq4} and reordering the terms will give the time derivative of angular components in the following canonical form
\begin{eqnarray}
    \frac{d\theta}{dt} &=& \kappa (-H_{\phi} + \alpha H_{\theta}),\\
    \frac{d\phi}{dt} &=& \kappa\dfrac{2}{\sin \theta} (-H_\theta + \alpha H_\phi),
\end{eqnarray} 
where $\kappa = \dfrac{\gamma}{1 + \alpha ^2}$.

  

\subsection{Inertial LLG equation in spherical coordinates}
Similarly, the iLLG equation can be written in spherical coordinates. As in the last section, vector multiplication of Eq.~\eqref{eq2} by \textbf{M} will give
\begin{equation} \label{eq5}
    \gamma \eta \tau M_{\textrm{s}}^{2} \frac{d^2\mathbf{M}_\perp}{dt^2}  = \gamma M_{s}^{2}\textbf{H}_\mathrm{eff \perp} - \gamma \eta M_\textrm{s}^{2}\frac{\mathbf{M}}{dt} + \textbf{M} \times \frac{\mathbf{M}}{dt}
\end{equation}
\noindent
where the subscript $\perp$ indicates the perpendicular component of \textbf{M}. Now the second derivative of the magnetization in spherical coordinates can be written as
\begin{widetext}
\begin{equation}
    \frac{d^2 \mathbf M}{dt^2} = -M_s \bigg(\frac{d\phi^2}{dt}  \sin ^2 \theta + \frac{d\theta^2}{dt}\bigg)\textbf{e}_\textrm{r} + M_s \bigg(\frac{d^2\phi}{dt^2}  - \frac{d\phi^2}{dt}  \sin \theta \cos \theta\bigg)\textbf{e}_\mathrm{\theta}+ M_s \bigg(2\frac{d\theta}{dt} \frac{d\phi}{dt}  \cos \theta + \frac{d^2\phi}{dt^2} \sin \theta\bigg)\textbf{e}_\mathrm{\phi}.
\end{equation}
\end{widetext}
\clearpage

Substituting this relation in Eq. \eqref{eq5} gives
\begin{align}
   \frac{d^2\theta}{dt^2} &= \frac{1}{\alpha \tau} \bigg(\alpha \tau \frac{d\phi^2}{dt}  + \gamma H_{\theta} - \alpha \frac{d\theta}{dt}  - \frac{d\phi}{dt} \sin \theta\bigg) \\
   \frac{d^2\phi}{dt^2} &=\frac{1}{\alpha \tau\sin \theta} \bigg(-2\alpha \tau \frac{d\theta}{dt} \frac{d\phi}{dt}  + \gamma H_{\phi} - \alpha \frac{d\phi}{dt} \sin \theta  \nonumber\\& - \frac{d\theta}{dt} \sin \theta\bigg)
\end{align}

The equilibrium condition in our simulation is defined as \{$\theta = \pi/2$,$\phi = \pi/2$, $d\theta/dt = 0$, $d\phi/dt = 0$\}. 


\section{\label{sec:level6}Magnetization dynamics in time domain}
\begin{figure}[t]
\centering
\includegraphics[width=\columnwidth]{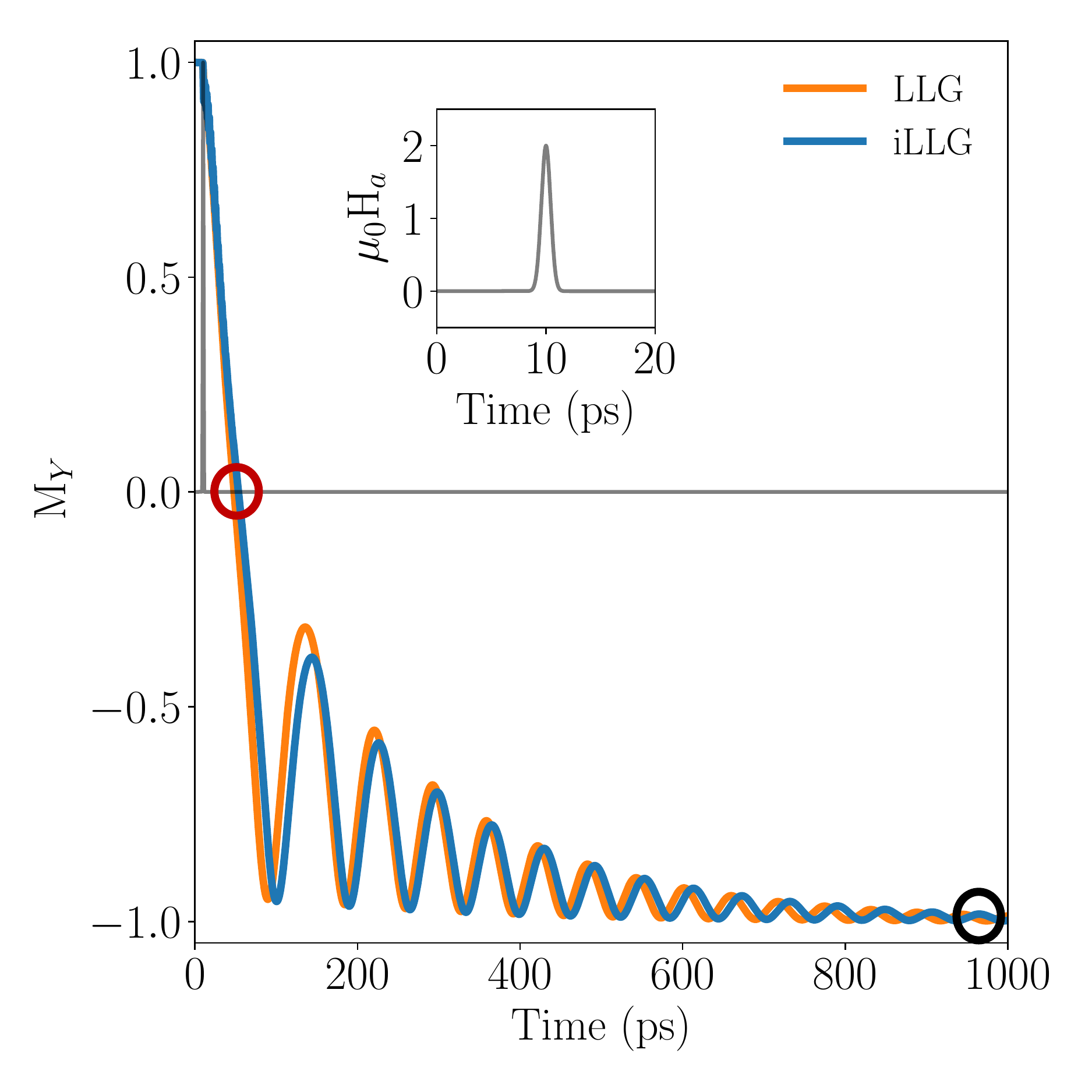}
\caption{\label{fig_5}Temporal evolution of the $y$-component of the magnetization for the LLG and iLLG dynamics for a case of precessional switching. The red circle indicates the point used to calculate the switching time, the black circle the time where the magnetization state (switched or not switched) was determined. The inset is the magnetic field pulse used to trigger the dynamics in the main figure.}
\end{figure}

Fig.~\ref{fig_5} shows the temporal evolution of the \textit{y} component of the magnetization when triggered by the magnetic field pulses represented in the inset. The black circular marker shows the point in time when the magnetization state is recorded to plot the diagram shown in Fig. \ref{fig_1}(c). The switching time in this study is defined as the time when the magnetization crosses the equator for the last time before reaching the steady state, red circle in Fig. \ref{fig_5}. In the following, the derivation of Eq. \eqref{en} in the main text is reported.

\section{\label{sec:energy} Energy diagram}
\begin{figure}[t]
\centering
\includegraphics[width=0.5\textwidth]{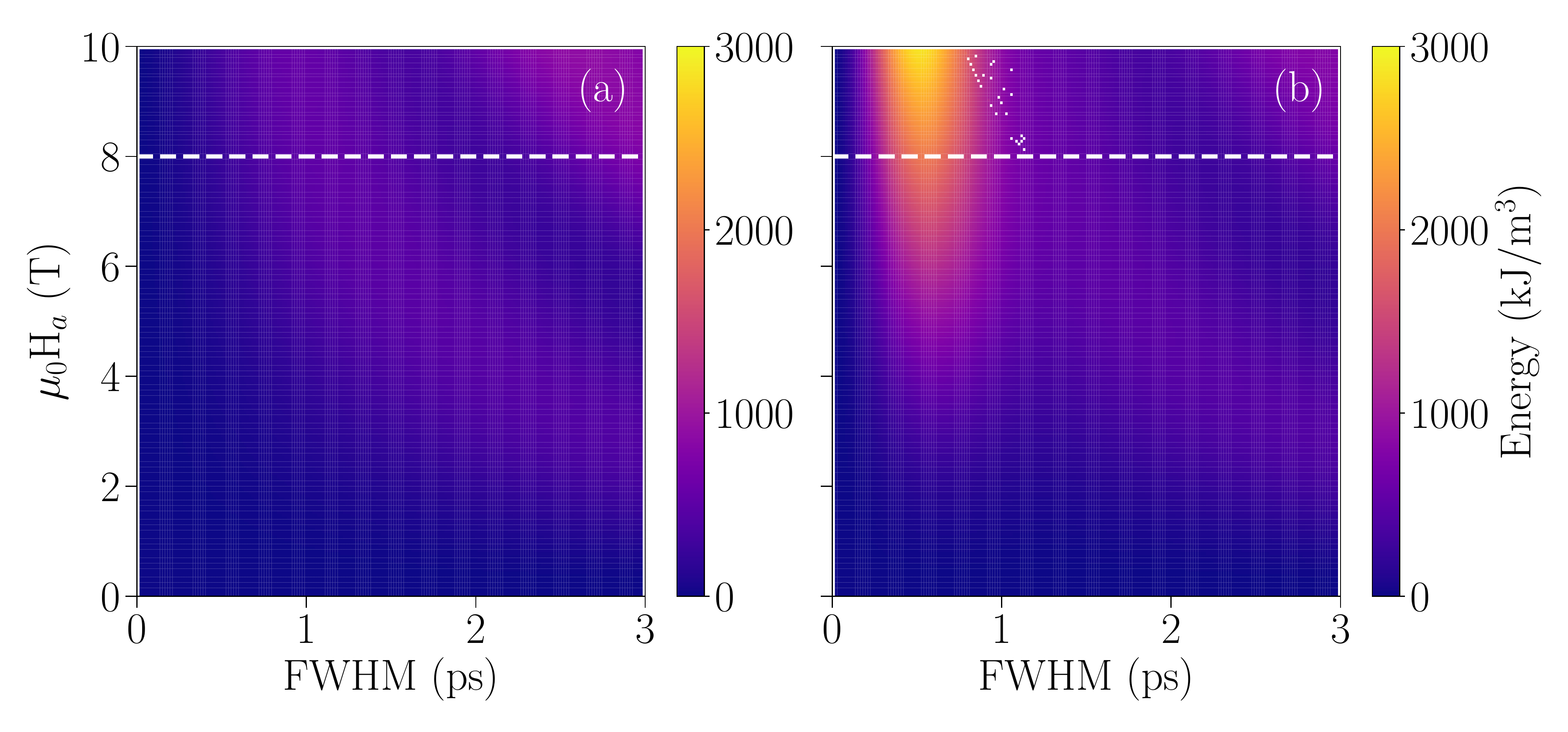}
\caption{\label{fig_6} Diagram of the energy density deposited by a magnetic field pulse as a function of its amplitude and width, calculated using (a) the LLG equation and (b) the iLLG equation. The horizontal white dashed line in correspondence of the 8 T value is the line cut plotted in Fig. \ref{fig_4} in the main text.}
\end{figure}

In Fig.~\ref{fig_6}, we plot the energy deposited by the pulse for the LLG and iLLG equations as a function of applied pulse width and amplitude. The energy is calculated using Eq. \eqref{en}, by taking a direct integral over a time duration of the full pulse width 6$\sigma$, with $\sigma\approx \text{FWHM}/2.4$. 

\subsection{Energy balance in iLLG dynamics}

For the sake of simplicity of notation, we consider the  iLLG dynamics in its dimensionless form:
\begin{equation}\label{eq:iLLG}
    \frac{d\bm m}{d t}=-\bm m \times \left(\bm h_\mathrm{eff}- \alpha\frac{d\bm m}{d t} - \xi \frac{d^2\bm m}{d t^2}\right) \,\,,
\end{equation}
where $\bm m(t)$ is the magnetization unit-vector (normalized by the saturation magnetization $M_s$), time is measured in units of $(\gamma M_s)^{-1}$, $\gamma=2.21\times 10^5$ A$^{-1}$s$^{-1}$m  is the absolute value of the gyromagnetic ratio, $\alpha$ is the dimensionless Gilbert damping parameter, and $\xi$ measures the strength of the inertial effects in the magnetization dynamics. The effective field is
\begin{equation}\label{eq:heff}
\bm h_\mathrm{eff}=-\frac{\partial g}{\partial \bm m} = -D\cdot \bm m + \bm h_a \,\,,
\end{equation}
which is expressed as the gradient of the free energy $g(\bm m, \bm h_a)=\frac{1}{2}\bm m\cdot D\cdot  \bm m -\bm h_a\cdot \bm m$, with $D=\mathrm{diag}(D_x,D_y,D_z)$ being the effective demagnetizing tensor (referred to principal axes and also including anisotropy) and $\bm h_a$ the external applied field. $\xi$ is related to the time scale of inertial dynamics and can be expressed as $\xi=\gamma M_s \alpha \tau$ to be consistent with Eq. (\ref{eq2}) and Ref. \cite{neeraj2021inertial}.

In order to investigate the energy aspects of the inertial magnetization dynamics, we dot-multiply Eq. \eqref{eq:iLLG} by the term in brackets, so that the right-hand side vanishes
\begin{equation}\label{eq:energy1}
    \left(\bm h_\mathrm{eff}- \alpha\frac{d\bm m}{d t} - \xi \frac{d^2\bm m}{d t^2}\right)\cdot \frac{d\bm m}{d t}=0  \,\,.
\end{equation}

By recalling Eq.~\eqref{eq:heff} and after some algebra, we obtain
\begin{equation}\label{eq:energy2}
    -\frac{d g}{d t}-\bm m\cdot \frac{d\bm h_a}{d t} - \alpha\left|\frac{d\bm m}{d t}\right|^2 -  
     \frac{d}{dt}\left(\frac{\xi}{2} \left|\frac{d\bm m}{d t} \right|^2 \right)=0  \,\,.
\end{equation}

Recasting terms, one arrives at the following equation for the energy balance in iLLG dynamics
\begin{equation}\label{eq:energy balance}
    \frac{d}{d t}\left(g + \frac{\xi}{2} \left|\frac{d\bm m}{d t} \right|^2 \right)=-\bm m\cdot \frac{d\bm h_a}{d t} - \alpha\left|\frac{d\bm m}{d t}\right|^2   \,\,.
\end{equation}
It is apparent that, for constant-in-time applied field, and when the Gilbert damping is zero ($\alpha=0$), the magnetization dynamics preserves the quantity:
\begin{equation}\label{eq:effective energy}
\tilde{g}(\bm m, d_t \bm m,\bm h_a)=g(\bm m,\bm h_a)+k(d_t \bm m ) \, ,
\end{equation}
where
\begin{align}
    k(d_t \bm m ) = \frac{\xi}{2} \left|\frac{d\bm m}{d t} \right|^2 \, ,
\end{align}
and where $d_t \bm m$  is a compact notation for $d \bm m /dt$. We note that the form of $\tilde{g}$ is analogous to the the total mechanical energy (potential + kinetic). For later use, we also introduce the following total energy defined by using, as potential energy, the Helmholtz free energy $a(\bm m)$:
\begin{equation}\label{eq:effective energy_a}
\tilde{a}(\bm m, d_t \bm m)=a(\bm m) + k(d_t \bm m ) \, ,
\end{equation}
where we recall that $g(\bm m,\bm h_a)=a(\bm m)-\bm m \cdot \bm h_a$.
On the basis of the the above considerations, we can conclude 
that Eq. \eqref{eq:iLLG} with $\alpha=0$ describes a conservative dynamical system
with energy equal to $\tilde{g}(\bm m, d_t \bm m, \bm h_a)$.

\subsection{Energy dissipation during magnetization switching for LLG and iLLG dynamics}

In order to compare the work to be done by the external field during magnetization switching in the presence or absence of inertia, one has to use Eq. \eqref{eq:energy balance}. Thus, the rate of change of the free energy $g$ during the switching process becomes
\begin{equation}\label{eq:rate of change}
   \frac{d \tilde{g}}{d t} = -\bm m\cdot \frac{d\bm h_a}{d t} - \alpha\left|\frac{d\bm m}{d t}\right|^2 \, .
\end{equation}

By using in Eq. \eqref{eq:rate of change} the expression of $g$  in terms of Helmholtz free energy, one can be easily convinced that Eq. \eqref{eq:rate of change} can be recast in a form which allows the interpretation of the involved terms in the framework of the laws of thermodynamics, i.e.
\begin{equation}\label{eq:rate of change2}
    \frac{da}{d t}= - \alpha\left|\frac{d\bm m}{d t}\right|^2 -\frac{d}{dt}\left(\frac{\xi}{2}  \left|\frac{d\bm m}{d t} \right|^2 \right)
    +\bm h_a\cdot \frac{d\bm m}{d t}  \,\,.
\end{equation}

Let us now consider the work performed by the field during a generic evolution of magnetization in the interval $[t_1,t_2]$.
By integrating Eq. \eqref{eq:rate of change2} with respect to time in that interval, we obtain
\begin{align}
   \Delta W =  \int_{t_1}^{t_2} \bm h_a\cdot \frac{d\bm m}{d t} dt = 
   \Delta a  + \left[ \frac{\xi}{2}  \left|\frac{d\bm m}{d t} \right|^2 \right]_{t_2}
    \nonumber \\ - \left[  \frac{\xi}{2}  \left|\frac{d\bm m}{d t} \right|^2\right]_{t_1}  
    +\int_{t_1}^{t_2}  \alpha\left|\frac{d\bm m}{d t}\right|^2 dt \,, \label{eq:rate of change3}
\end{align}
where 
\begin{align}
\label{eq:rate of change4}
    \Delta {a} =  a(\bm m(t_2)) - a (\bm m(t_1)) \, ,
\end{align}
and where we have denoted with $\Delta W$ the work performed on the magnetic particle during the process.

In computing $\Delta W$ for a switching process, it is convenient to use the initial time instant $t_0$, i.e. the instant before which the applied magnetic field is zero. At $t_0$ the magnetization is in an equilibrium point $\bm m_0$, which corresponds to a minimum of the Helmholtz free energy $a(\bm m)$ and $d \bm m/ dt =0 $. Now, if $t_{p}$ is the time instant when the external field pulse is again zero after reaching a finite value, one could choose $t_2=t_p$, and compute $\Delta W$ as
\begin{align}
    \Delta W = \int_{t_0}^{t_p} \bm h_a\cdot \frac{d\bm m}{d t} dt  
    \, .
\end{align}

This computation has the difficulty that from Eq. \eqref{eq:rate of change3} one is required to know the values of $\bm m$ and $d \bm m /dt $ in both $t_0$ and $t_p$. A simplified formula can be obtained by taking into account that
\begin{align}
\label{eq:rate of change5}
    \Delta W = \int_{t_0}^{t_p} \bm h_a\cdot \frac{d\bm m}{d t} dt  
    \approx \int_{t_0}^{\infty} \bm h_a\cdot \frac{d\bm m}{d t} dt 
    \, ,
\end{align}
assuming that the applied field is negligible for $t>t_p$, which is a good approximation for a Gaussian-like magnetic field pulse if one takes $t_p\approx 6\sigma$. In addition, in the typical switching process, the asymptotic magnetization state ($t \rightarrow \infty$) and the initial  magnetization state are rest states (with $d \bm m /dt =0$) and they have the same value of the Helmholtz free energy. By using this fact in Eq. \eqref{eq:rate of change3} and Eq. \eqref{eq:rate of change5}, we arrive to the equation
\begin{align}
\label{eq:work}
    \Delta W =  \int_{t_0}^{t_p} \bm h_a\cdot \frac{d\bm m}{d t} dt
    \approx  \int_{t_0}^{\infty}  \alpha\left|\frac{d\bm m}{d t}\right|^2 dt \, 
    \, .
\end{align}
Eq. \eqref{eq:work} is the dimensionless version of Eq. \eqref{en} used for computations in this paper.

An analogous result can be obtained in cyclic transformations starting and ending in the same magnetization state (e.g. as in the case of an hysteresis loop). In this case we obtain
\begin{align}
    0&=-\oint \left[ \alpha\left|\frac{d\bm m}{d t}\right|^2 +
    \frac{d}{dt}\left(\frac{\xi}{2}  \left|\frac{d\bm m}{d t} \right|^2 \right)\right] d t +  \oint \bm h_a\cdot \frac{d\bm m}{d t} dt \,, \label{eq:dissipated energy1}\\
\Delta W &= \oint \alpha\left|\frac{d\bm m}{d t}\right|^2 d t \geq 0 \,\,, \label{eq:dissipated energy3}
\end{align}
The right hand side represents the amount of energy dissipated into heat through intrinsic entropy production.

\end{document}